# A COST-EFFECTIVE RAPID-CYCLING SYNCHROTRON

SERGEI NAGAITSEV[1]

*Fermilab, Batavia, IL 60510, USA*
*nsergei@fnal.gov*

VALERI LEBEDEV

*Fermilab, Batavia, IL 60510, USA*
*val@fnal.gov*

The present Fermilab proton Booster is an early example of a rapidly-cycling synchrotron (RCS). Build in 1960s, it features a design in which the combined-function dipole magnets serve as vacuum chambers. Such a design is quite cost-effective, and it does not have the limitations associated with the eddy currents in a metallic vacuum chamber. However, an important drawback of that design is a high impedance, as seen by a beam, because of the magnet laminations. More recent RCS designs (e.g. J-PARC) employ large and complex ceramic vacuum chambers in order to mitigate the eddy-current effects and to shield the beam from the magnet laminations. Such a design, albeit very successful, is quite costly because it requires large-bore magnets and large-bore rf cavities. In this article, we will consider an RCS concept with a thin-wall metallic vacuum chamber as a compromise between the chamber-less Fermilab Booster design and the large-bore design with ceramic chambers.

---

[1]Also: Department of Physics, The University of Chicago, Chicago, IL 60637, USA

# I Introduction

Fermilab's Proton Improvement Plan II, or PIP-II, will enable the world's most intense neutrino beam and help scientists search for rare particle physics processes [1]. The PIP-II goal is to deliver 1.2 MW of proton beam power from the Fermilab Main Injector, over the energy range 60 – 120 GeV, at the start of operation of the LBNF/DUNE program. PIP-II provides a variety of upgrade paths to higher beam power from the Main Injector, as demanded by the neutrino science program. Delivery of more than 2 MW to the LBNF target in the future will require replacement of the existing Booster. The most straightforward strategy would be to extend the 0.8-GeV PIP-II linac to 1.5-2 GeV and to inject at this energy into a new 8-GeV rapid cycling synchrotron (RCS). In this article, we will describe a cost-effective concept of a 2-8 GeV proton RCS, based on Ref [2] and the design with a primary goal to support the Fermilab Main Injector operation with 2 MW beam power on the neutrino target. For this concept, we will assume a 2-GeV, 2-mA H$^-$ linac as an injector, capable of the bunch-by-bunch chopping for the longitudinal painting at injection into the RCS. We will also assume a fixed-energy (8 GeV) accumulation ring (the existing Fermilab Recycler ring) to collect several RCS batches during the Main Injector 8-120 GeV ramp. We assume that the Recycler and the Main Injector have equal circumferences, thus, allowing for a single-turn bucket-to-bucket transfer. Figure 1 schematically shows the proposed Fermilab site layout.

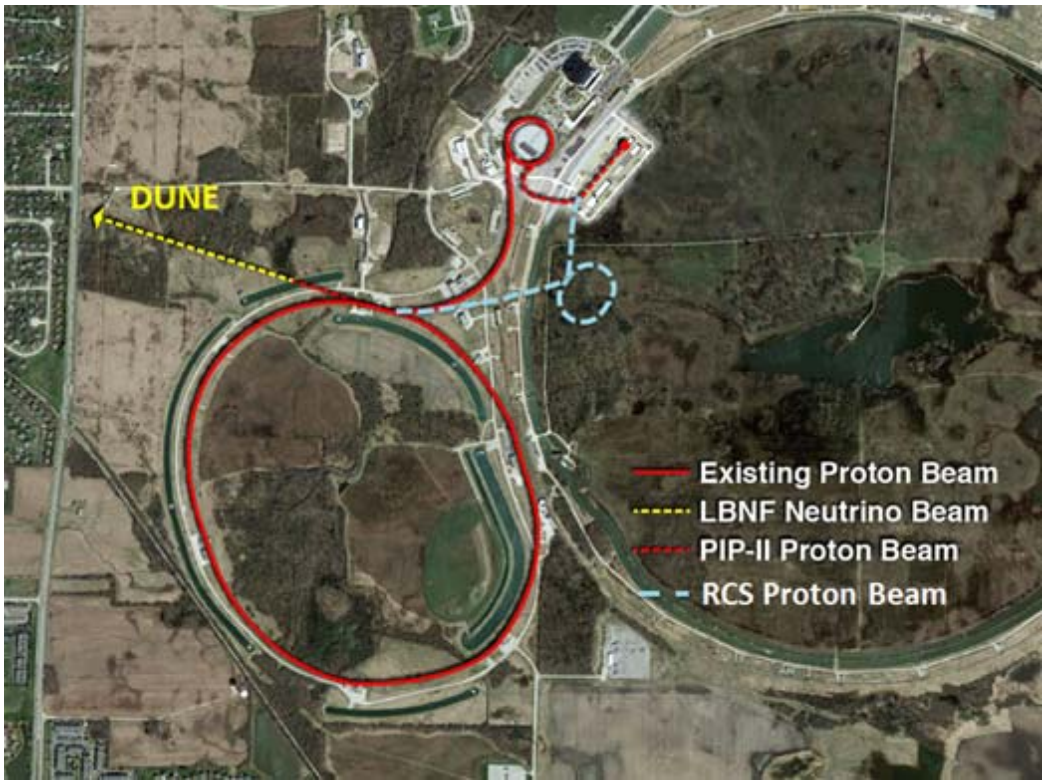

Figure 1: The schematic RCS site layout.

# II Main RCS Parameters

To support the 2-MW operation of the Fermilab Main Injector (MI) for the neutrino science program, a rapid cycling synchrotron (RCS) must deliver 1.6·10$^{14}$ protons to the MI each cycle, which can vary in length from ~0.7 sec (for the 60 GeV extraction energy) to ~1.2 sec (120 GeV). The RCS has a shorter circumference than the Recycler and therefore several RCS cycles are required to fill it. Balancing the impacts of beam space charge at injection, instabilities, the magnetic field strength, and the repetition rate, the circumference is chosen to be 1/6 of the MI circumference and the repetition rate is chosen to be 10 Hz. The main parameters of the RCS are presented in Table I.

**Table I: Main Parameters of RCS**

| Energy, min/max, GeV | 2/8 |
|---|---|
| Repetition rate, Hz | 10 |
| Circumference, m (MI/6) | 553.2 |
| Tunes, $\nu_x/\nu_y$ | 18.42 / 18.44 |
| Transition energy (kinetic), GeV | 13.3 |
| Number of particles | 2.6 x $10^{13}$ |
| Beam current at injection, A | 2.2 |
| Transverse 95% normalized emittance at injection, mm mrad | 22 [2] |
| Space charge tune shift, at injection | 0.07 [3] |
| Normalized acceptance at injection, mm mrad | 40 |
| Harmonic number for main RF system, $h$ | 98 |
| Harmonic number for 2-nd harmonic RF system, | 196 |
| RF bucket size at injection, eV-s | 0.38 |
| Injection time for 2-mA linac current, ms | 2.1 |
| Total beam power from linac, kW | 90 [4] |
| Total beam power delivered by RCS, kW | 340 |

The requirements for reliable and efficient operations of the RCS lead to some specific design decisions. To avoid transition crossing, the transition energy is chosen to be outside of the machine energy range. The focusing lattice needs to be relatively insensitive to focusing errors and synchro-betatron resonances. The latter is achieved by having zero dispersion in the straight sections, where RF cavities are located. A FODO lattice with a missing dipole for dispersion suppression is a natural choice, resulting in modest requirements for magnets and vacuum chambers as well as strongly suppressed nonlinear resonances. However, to avoid the stripping foil overheating by the injected beam (see Section 4), the beta-functions at the foil location need to be increased relative to their nominal values of the FODO structure. Therefore, the optics for seven half-cells near the injection point was modified. This resulted in an increase of the geometric mean of beta-functions at foil, $\sqrt{\beta_x \beta_y}$, from 5.5 m to 20.5 m with the corresponding decrease of foil heating by more than an order of magnitude. Note, that if such large beta-function values are used in the arcs, it would result in the transition energy being within the RCS energy range regardless of what kind of single cell optics is used (FODO, doublet, triplet, etc.). It would also result in an increase of beam pipe size and its eddy-current heating.

The ring is designed as a racetrack (two long straight sections and two 180° arcs) with the same distance between the centers of quads through the entire ring with exception of the injection region. One long straight section is for the RF cavities and the other one is used for injection, extraction, and beam collimation. The F and D quads have the same focusing strength and are connected serially with the dipoles. Eight quadrupoles, four in the injection and four in the extraction regions, have a larger aperture and length, but they have the same integral strength. The tune and optics corrections are performed via additional corrector coils wound in each quadrupole. Figure 2 presents the beta-functions, the dispersion, and the beam envelopes for the half of the ring. The injection region is shown on the left side of the plot, where two quadrupole doublets replace six quadrupoles of the FODO structure. The rest of the ring, including another straight line not shown in the picture, has the regular FODO structure. The betatron phase advance per cell is 102°. The strong focusing results in small beam sizes and small dispersion. That, in its turn, results in a small synchrotron beam size and, consequently, a small difference between horizontal and vertical beam envelopes through the entire ring.

---

[2] A 12% emittance dilution is assumed during acceleration and transfer to Recycler, where the design value for the emittance is 25 mm mrad.

[3] This value is estimated for a KV-like transverse distribution and the longitudinal bunching factor of 2.2 that are obtained by the beam painting as presented below in Section 5.

[4] We imply here a 4% loss of the injected beam. See details below.

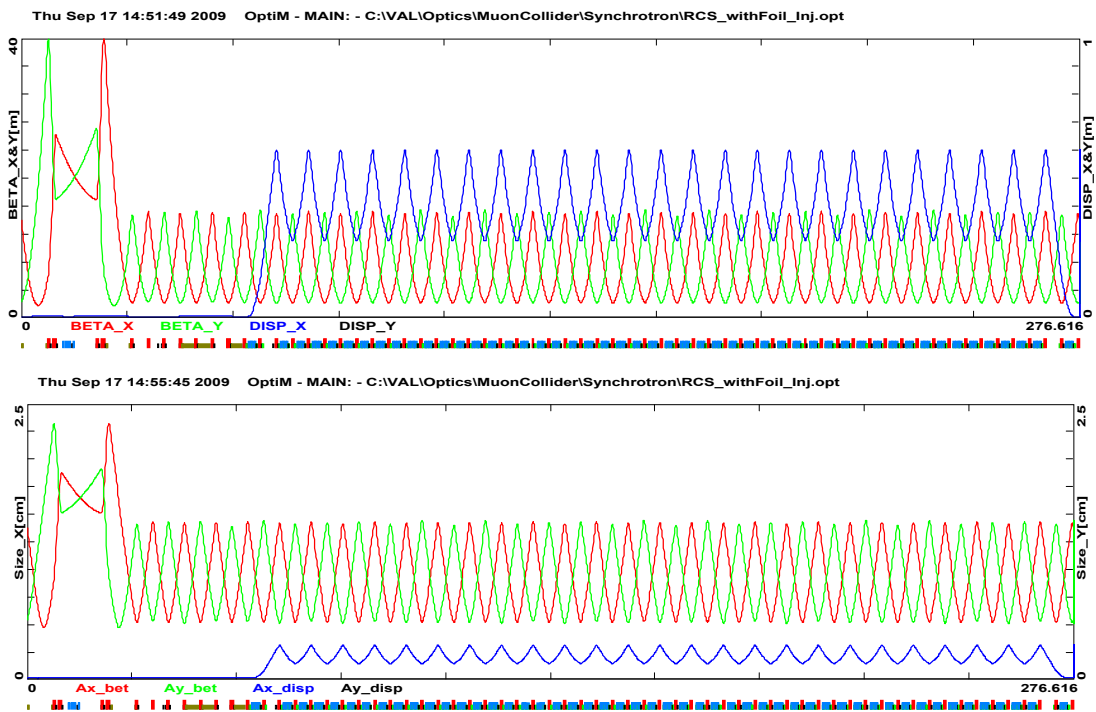

Figure 2: The beta-functions and dispersion (top) and the beam envelopes (bottom) for a half of the ring (the injection & extraction straight and downstream arc.) The beam envelopes are shown for $\varepsilon_n$=40 mm mrad, $E_k$ = 2 GeV, $\Delta p/p$ = 5 x 10-3.

Table II presents the structure of the machine period. The FODO structure is built so that for a perfectly periodic lattice it would have 66 cells with two 7-cell long straight sections. Considering that 7 half-cells of the injection straight are used for injection optics and they have 4 quads instead of 6, one obtains that altogether there are 130 quadrupoles and 100 dipoles. To ease the magnet power supply voltage requirements, the dipoles and the quadrupoles of all cells are included into a 10-Hz resonance circuit. Every quadrupole has an associated corrector package (designated as sF and sD in Table II). The general corrector package contains a trim dipole coil (horizontal near the F quads and vertical near the D quads) and a sextupole coil at the locations with non-zero dispersion. Table II presents the required strength of the trim magnets.

**Table II: The structure of a periodicity element for the RCS at 8-GeV kinetic energy**

| Name | S[cm] | L[cm] | B[kG] | G[kG/cm] | S[kG/cm/cm] |
|---|---|---|---|---|---|
| qF | 65.9 | 65.9 | 0 | 1.7675 | 0 |
| o2 | 85.9 | 20 | | | |
| sF | 105.9 | 20 | 0 | 0 | 0.185 |
| o1 | 135.9 | 30 | | | |
| bD | 349.116 | 213.216 | 8.7375 | 0 | 0 |
| o | 419.116 | 70 | | | |
| qD | 485.016 | 65.9 | 0 | -1.7634 | 0 |
| o2 | 505.016 | 20 | | | |
| sD | 525.016 | 20 | 0 | 0 | -0.324 |
| o1 | 555.016 | 30 | | | |
| bD | 768.232 | 213.216 | 8.7375 | 0 | 0 |
| o | 838.232 | 70 | | | |

Two groups of dipole correctors perform the orbit correction: the regular correctors and the injection correctors. There are altogether 124 regular dipole correctors: 62 horizontal and 62 vertical correctors located near the focusing and the defocusing quads, respectively. Their strength is determined by the accuracy of dipole fields (<0.3%) and the quadrupole alignment (<1 mm) resulting in the strength of 11 kG-cm. The corrector ramp rate is determined by the duration of the cycle and was chosen to be 1 kG-cm/ms. The regular dipole correctors are located in the cell locations from 8 to 131. There are four injection correctors on each plane independently controlling the beam position and angle on the stripping foil. They are located on both sides of the injection straight. The injection correctors perform the linac beam painting into transverse aperture of the ring (see below). They have a larger strength (30 kG-cm) and a much higher ramping rate (30 kG-cm/ms). There is also a permanent orbit bump in the extraction region. The bump is introduced by a vertical displacement of three quadrupoles and therefore does not require special correctors.

To ease the betatron tune correction, the trim coils in quadrupoles are split into two large quadrupole families, F and D, which do not include trim coils in 36 quadrupoles assigned for optics correction. Each F and D family includes coils in 47 quads and is powered from one power supply. The optics correction quads are located on the both ends of each long straight, and in the middle and on the both ends of each arc. Thus, the optics correction quads are: 3-6, 11-14, 16, 18, 37-42, 62, 64, 66-69, 77-80, 82, 84, 103-108, 128 and 130. The strength of the trim quadrupoles is determined by (1) the focusing errors, determined by finite accuracy of the quadrupole fabrication ($\sim 2 \cdot 10^{-3}$), (2) the different dependence of magnet strength on the power supply current for dipoles and quadrupoles due to different magnetization of their cores ($\leq 5 \cdot 10^{-4}$), (3) a different field correction for dipoles and quads excited by the eddy currents in the vacuum chamber ($5 \cdot 10^{-4}$), and (4) a sufficiently large tune correction span. For the F and D families, the correction value of $\pm 1\%$ ($\int GdL = 1.1$ kG) was chosen resulting the tune correction range of $\pm 0.25$ for both planes. The optics correction trim quads have a factor of two higher strength of 2.2 kG. Note that all large aperture quads belong to the group of optics correction quads and, consequently, have correcting coils.

There are 98 sextupoles in the ring. They are located at the positions 16-64 and 82-130 and are split into two families F (50 sextupoles) and D (48 sextupoles). The sextupole strengths required to correct the natural machine chromaticities, $\xi_x \approx \xi_y \approx -25$, to zero are $\int S_F dL = 3.7$ kG/cm and $\int S_D dL = -6.8$ kG/cm at the beam energy of 8 GeV. Considering that the RCS will be operated with chromaticities in the ranges of -10 to -20 we chose the maximum strength to be 4 kG/cm. This still leaves a sufficiently large margin between the operational and maximum strengths of sextupoles.

To minimize the vertical dispersion and *x-y* coupling, twelve skew-quadrupole correctors are installed in the positions 3, 15, 16, 64, 65, 66, 67, 81, 82, 130, 131 and 132. These correctors have the length of 20 cm, and the integral strength of 5 kG.

The relatively low current of the injection linac, 2 mA, results in long injection time, 2.1 ms or ~1000 turns. To minimize the required beam energy correction, the injection begins 1.05 ms before the magnetic field reaches its minimum (assuming the 10-Hz RCS cycle). The corresponding variation of the bending field of ±0.11% is compensated by dipole correctors.

## III  RCS Vacuum Chamber

The vacuum chamber for the RCS needs to satisfy a set of opposing requirements. To minimize the effects of eddy currents in a conductive beam pipe, a thin-wall small-radius pipe would be optimum. To maximize the mechanical stability, a thick-wall pipe would be optimum. To minimize the transverse impedance, a high-conductivity wall with a large-radius pipe would be optimum. In summary, the competing effects are:

- the shielding and distortion of the dipole bending field by the eddy currents, excited in the vacuum chamber;
- the vacuum chamber mechanical integrity under the atmospheric pressure;
- the vacuum chamber heating by the eddy currents;
- the transverse impedance due to the wall resistivity;
- the ring acceptance.

The compromise resulted in a round stainless-steel vacuum chamber with the external radius of 22 mm and the wall thickness of 0.7 mm. Details are discussed below.

The complex amplitude of magnetic field produced by eddy currents, excited in the vacuum chamber of radius $a_w$ with the wall thickness $d_w$ and the wall conductivity $\sigma_w$ by the AC component of dipole magnetic field alternating with frequency $\omega_{ramp}$ is:

$$\Delta B_y(0, y) = iB_{AC}\left(1 + \frac{\pi^2}{12} + \frac{\pi^4}{240}\frac{y^2}{a_w^2} + ...\right)\frac{a_w d_w}{\delta_w^2}, \quad \delta_w = \frac{c}{\sqrt{2\pi\sigma_w\omega_{ramp}}} << a_w \ . \tag{1}$$

One can see that this field correction is shifted by 90 deg. relative to the AC component of the dipole magnetic field. The first addend in the parenthesis is related to the eddy currents excited in the wall by bending field and the other two addends are related to the multiple reflections of this current in the poles of a dipole. For the magnetic field changing as $B(t) = B_{DC} - B_{AC}\cos(\omega_{ramp}t)$, the last addend corresponds to the sextupole field with the sextupole strength equal to:

$$S(t) = B_{AC}\frac{\pi^4}{120}\frac{d_w}{a_w\delta_w^2}\sin(\omega_{ramp}t) \ . \tag{2}$$

The relative value of dipole correction of the field (sum of the first two addends in Eq. (1)) is equal to zero at the cycle start and end. It achieves its maximum of $|\Delta B/B|=8.5\cdot10^{-4}$ at 16 ms within the acceleration cycle. Similarly, in the case of a changing quadrupole field in a quadrupole magnet there is a quadrupole field correction with a relative value approximately a half of the dipole correction. Consequently, keeping constant tunes during the acceleration cycle requires a quad current correction $\Delta I/I \approx 4.3 \cdot 10^{-4}$. As for the field correction in a dipole, the sextupole field correction is zero at the cycle beginning and end. At 16 ms it results in the maximum contribution to the machine chromaticity: $\Delta \xi_x \approx 1.03$ and $\Delta \xi_y \approx -0.85$. These values are a small fraction of the natural machine chromaticity and can be easily compensated by the machine sextupoles. Note that the sign of the sextupole correction is "focusing" in the acceleration part of the cycle and "defocusing" at the deceleration part. Tracking studies show that if only the machine sextupoles and the eddy current sextupoles are considered, the dynamic aperture exceeds the machine aperture by about a factor of 4.

There must be a sufficiently large safety margin for the mechanical stresses of vacuum chamber to ensure the reliable operation for the long lifetime of the machine. For a perfectly round vacuum chamber, the stress in the material due to atmospheric pressure, $P_{atm}$, is determined as

$$\sigma_{cmpr} = P_{atm} \frac{a_w}{d_w}, \qquad (3)$$

and is equal to 3.1 N/mm². If the chamber is slightly elliptic there is an additional bending stress, equal to

$$\sigma_{bend} = \frac{9}{4} P_{atm} \frac{\Delta a_w}{a_w} \left( \frac{a_w}{d_w} \right)^2, \qquad (4)$$

where $\Delta a_w$ determines the ellipse semi-axes to be equal to $a_w \pm \Delta a_w$. Assuming a comparatively conservative ellipticity of the chamber $\Delta a_w / a_w = 0.02$, corresponding to $a - b = 0.88$ mm, we obtain $\sigma_{bend} = 8.9$ N/mm². The sum of these two stresses is equal to 12 N/mm². It is ~20 times smaller than the maximum acceptable stress (the yield stress) for the stainless steel of ~200 N/mm². A further reduction of vacuum chamber thickness is still possible but a thickness below 0.5 mm would jeopardize the long-term stability and the integrity of the chamber.

The eddy currents produce the vacuum chamber heating load because of ohmic losses. The thermal power per unit length is

$$\frac{dP}{dz} = \frac{\pi \sigma_R d_w a_w^3 \omega_{ramp}^2}{c^2} B_{AC}^2. \qquad (5)$$

For the nominal RCS parameters this heating load is equal to ~10 W/m. This power level does not require water or forced air cooling. A conservative air cooling estimate for the case of convective cooling is based on the heat transfer coefficient of $10^{-3}$ W/cm²/K. If one neglects the thermal conductivity of the chamber it results in a temperature increase of 15 C (compared to ambient) on the both sides of the chamber, where the current is concentrated. Further increase of the ramp frequency or the AC component of magnetic field would require forced air cooling.

If there were no thermal effects associated with the eddy currents, the non-linear eddy current fields of a metallic chamber could be passively compensated with correction windings on the vacuum chamber wall like it was implemented at the Brookhaven AGS Booster [3]. The scheme successfully corrects the non-linear field for arbitrary ramp rates and allows for reasonably thick-walled, low impedance and possibly large aperture vacuum chambers.

## III.1 Vacuum system

The RCS is composed of 132 half cells. The vacuum layout for a half-cell is shown schematically in Figure 3. The required vacuum level in this accelerator is not stringent: $\leq 10^{-7}$ Torr. However, this system does require some novel features such as RF shielding on ports and bellows (to reduce the ring impedance). Experiences from existing accelerators will provide the basis for a design and a cost estimate.

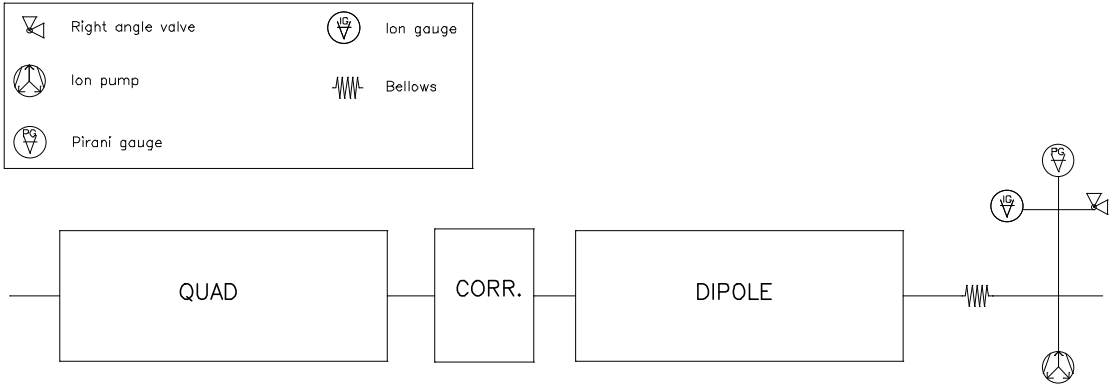

Figure 3: Typical vacuum layout for an RCS half-cell.

The required beam vacuum tube size for the quadrupoles and the dipoles is 44 mm OD, transitioning to a standard 2-inch OD tube for the rest of the vacuum chamber. The wall thickness is 0.7 mm inside the magnets. It can be larger at other places. Bellows and pumping ports will require RF shielding. In order to minimize the secondary emission yield from the chamber wall, the entire vacuum chamber will be coated with TiN. This has been done successfully at the SNS and the PEP II B-Factory.

Because of magnetic permeability requirements, the vacuum chamber will be constructed of the 316LN stainless steel. It will be electropolished and hydrogen degassed prior to coating with TiN. Alternatively, Inconel 718 is being considered as a promising choice of material for the quadrupole and the dipole beam tubes. Compared to the stainless steel, it has higher electrical resistivity to reduce the eddy current effects and higher strength for structural stability. Disadvantages include the high cost and the challenge of coating with a thin and highly uniform copper layer to reduce impedances. It would not be required for the 316LN stainless-steel option.

Vacuum levels will be maintained with ion pumping. Additional pumping may be needed in some areas depending upon out-gassing rates of specific accelerator components or special needs in certain areas (interfaces between accelerators, diagnostic components, etc.). The system will not require an *in-situ* baking. Ion gauges and convection gauges will provide vacuum read back.

## III.2 RCS Impedance

Let us now estimate the impedance of the vacuum chamber. At low frequencies ($f \leq 0.5$ GHz), the transverse impedance is dominated by the wall resistivity. For a thin-wall chamber and the frequency range corresponding to the lowest betatron sidebands, the real part of the impedance per unit length can be approximated by the following equation:

$$Z_\perp(\omega) = Z_0 \frac{c^2}{4\pi^2 \sigma_R \omega a_w^3 d_w} \quad , \quad \delta_w = \frac{c}{\sqrt{2\pi\sigma_w \omega}}, \quad \sqrt{a_w d_w} \geq \delta_w \geq d_w \quad , \tag{6}$$

where $Z_0 \approx 377\ \Omega$ is the impedance of vacuum. Comparing Eqs. (5) and (6) one can see that the transverse impedance and the vacuum chamber heat load are closely related:

$$Z_\perp(\omega)\frac{dP}{dz} = \frac{Z_0}{4\pi}\frac{\omega_{ramp}^2}{\omega}B_{AC}^2 \ . \tag{7}$$

Their product does not depend on the vacuum chamber radius, thickness and material. Varying the parameters to reduce the heating term results in an increase of the impedance and vice versa. Figure 4 compares impedances of the nominal stainless-steel vacuum chamber and the vacuum chamber similar to the Fermilab Booster (where the poles of laminated dipole represent also the vacuum chamber wall). The high value of μ and the presence magnet laminations greatly increase the impedance at high frequencies. In particular, there is a large difference for frequency above 50 MHz, where the active transverse damping is very difficult. Note that the calculations of the transverse impedance of the steel poles assume μ=500 over the entire frequency range. This assumption is not justified for frequencies above ~10 MHz where the skin depth becomes comparable to the domain size. Consequently, the impedance is somewhat smaller at very high frequencies but is still well above the impedance of the stainless-steel chamber. Figure 5 presents the betatron tune shifts due to the beam interaction with the stainless-steel vacuum chamber in the approximation of a zero betatron tune spread. The growth time for the lowest betatron sideband is about 150 turns. A bunch-by-bunch transverse damper can suppress this instability easily. Instabilities at frequencies corresponding to intra-bunch motion ($f > 50$ MHz) are damped by the chromaticity (see below).

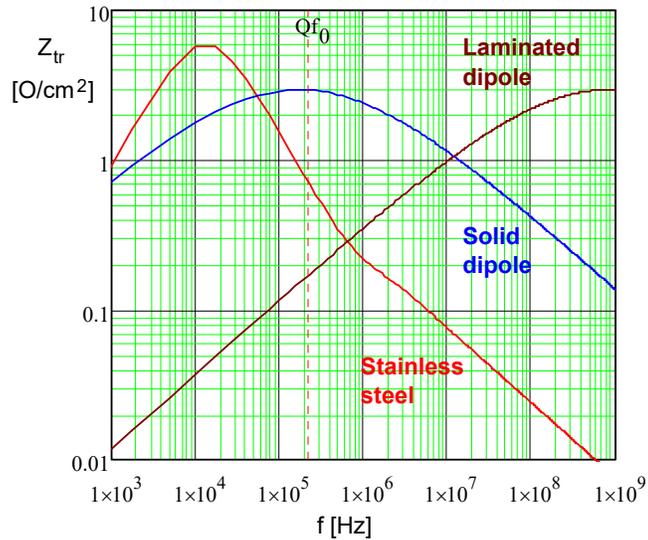

Figure 4: Dependence of the transverse impedance per unit length on frequency; red line – a thin stainless-steel vacuum chamber with radius 22 mm and wall thickness of 0.7 mm; blue and brown lines – vacuum chamber walls are poles of solid and laminated dipoles correspondingly with a gap of 44 mm and μ=500, the lamination thickness of 0.63 mm.

From the above discussion, there are many reasons to keep the vacuum chamber size being sufficiently small. It also shows that a ceramic vacuum chamber is not really beneficial (while significantly more expensive): a reduction of conducting layer thickness could reduce the vacuum chamber heating but it results in an increase of transverse impedance being the same as for a stainless-steel chamber. If the vacuum chamber wall is too thin or absent as in the Fermilab Booster, the beam interacts with the steel laminations of magnets resulting in much higher impedances and instability growth rates.

There is a sufficiently large margin between the 95% normalized emittance of 25 mm-mrad and the beam boundary set at 40 mm-mrad. At injection the beam boundary corresponds to the maximum beam size of 14 mm leaving 6 mm for orbit distortions in the both planes. This value is within normal operational orbit distortions of the present Fermilab Booster. The sagitta in the dipole is equal to 1.7 cm. To avoid aperture loss, the vacuum chamber in the dipoles must be bent to follow the beam bending with 33.9-m radius.

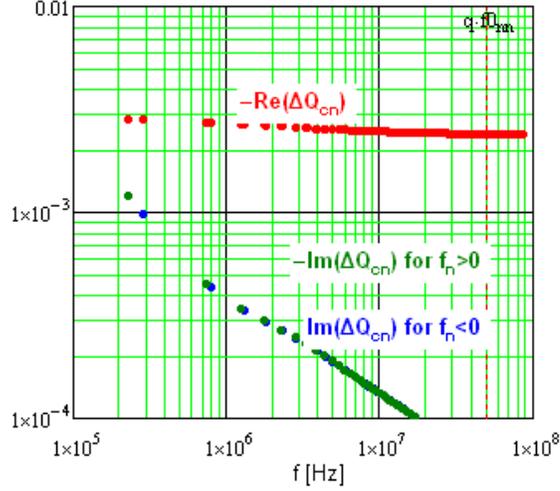

Figure 5: Real and imaginary tune shifts for the stainless-steel vacuum chamber; $a_w$=22 mm, $d_w$=0.7 mm.

### III.3 Beam Stability

A beam is stable, if for its every coherent mode a sum of Landau damping and the damper-induced rates exceed the impedance-associated growth rate. Landau damping is extremely sensitive to a ratio of the effective space charge tune shift to the synchrotron tune [4]. For the discussed parameters, this ratio is $q$=2.4 for a 3D Gaussian beam, and it is $q$=1.6 for KV model at injection energy. Since the space charge parameter $q$ is not extremely high, the strong space charge theory [4] can be used only as a rather rough approximation. According to that, the $0^{th}$ head-tail mode does not have any Landau damping at all, while the first mode has rather fast damping rate $\Lambda_1 \approx (0.2-0.4)Q_s$, and higher modes should not be seen at all.

The coherent growth rate can be estimated by the air-bag model (Ref. [5], Eq. (6.188)):

$$\mathrm{Im}\,\Omega = \frac{Nr_0 c}{2\gamma T_0^2 \omega_b} \sum_{p=-\infty}^{\infty} Z_\perp(\omega') J_l^2(\omega' \hat{z}/c - \chi) \ , \quad \omega' = pM\omega_0 + \mu\omega_0 + \omega_b \ . \qquad (8)$$

For the resistive wall impedance, given by Eq. (6), it results in a growth rate equal to:

$$\mathrm{Im}\,\Omega/\omega_0 = \frac{Nr_0\delta_0\overline{\beta}_x}{2\pi\gamma a^3}\Gamma(l,\chi,\mu);$$

$$\delta_0 = c/\sqrt{2\pi\sigma\omega_0}\,; \tag{9}$$

$$\Gamma(l,\chi,\mu) \equiv \sum_{p=-\infty}^{\infty}\sqrt{\frac{\omega_0}{|\omega'|}}J_l^2\left(\omega'\hat{z}/c - \chi\right)\mathrm{sgn}(\omega')$$

The most unstable coupled-bunch mode for the betatron tune $\nu_b$=18.44 is the mode $\mu$ = -19. The mode-factors $\Gamma$ for this coupled-bunch number are presented in Fig. 6. Without a damper, the $0^{th}$ head-tail mode can be only stabilized for the head-tail phase $\chi$. Assuming $\chi$=1.5, the growth rate for the $1^{st}$ head-tail mode is calculated to be Im($\Omega$)/$\omega_0$=0.01$Q_s$, which is 20-40 times smaller than the Landau damping rate for this mode. Thus, there is a significant safety factor for beam coherent stability.

For the extraction energy, the synchrotron frequency goes fast to its minimal value at the very end of the cycle. This results in $q$=10 at extraction, making the first 3-4 head-tail modes formally unstable. However, the growth times of these modes are calculated as 20 ms or larger and are too long to cause concern, since the entire acceleration time is 50 ms.

The longitudinal microwave stability threshold (Keil-Schnell) is calculated as 10-20 times above the nominal beam current (at all energies), so this instability should be no problem.

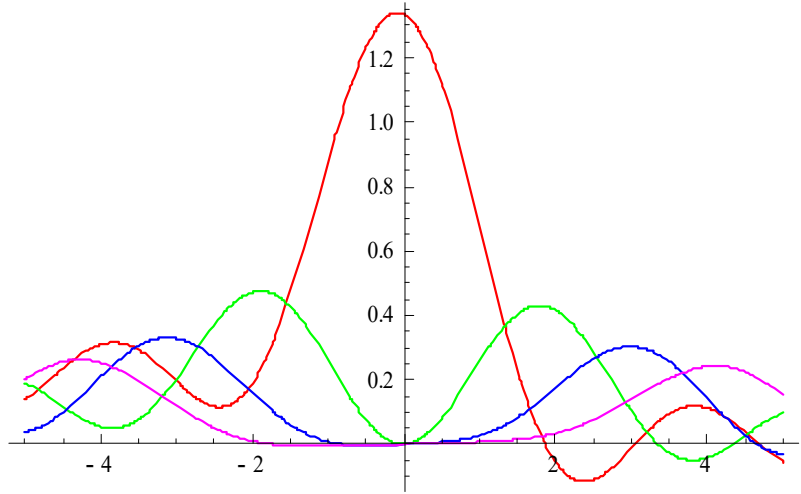

Figure 6: Mode factors $\Gamma$ as functions of the head-tail phase $\chi$ for the head-tail modes 0 (red), 1 (green), 2 (blue) and 3 (magenta), and the coupled-bunch mode number -19.

## IV  RCS RF Systems

Figure 7 presents time dependences of beam and RF system parameters during the acceleration and Table III shows the main parameters of the RF system. To reduce the betatron tune shifts due to the beam space charge at injection and through acceleration, we plan to use a double harmonic RF system so that the voltages of the fundamental (first harmonic) RF system and an additional RF system operating at its second harmonic could create a longitudinal potential well with a flat bottom through the entire accelerating cycle (see below). Presently, we plan to inject the beam into the ring with the flat-bottom potential well and, then, gradually reduce the second harmonic RF voltage in the second half of the accelerating cycle; so that the beam would be matched to the Recycler RF which is not planned to have the second harmonic RF on at injection. The match also implies that the longitudinal emittance needs to be increased from 0.35 to 0.4 eV s (or if necessary even to 0.6 eV s) during the acceleration. It can be achieved by manipulations with the longitudinal quadrupole damper similarly to how it is done now in the Booster. However, if necessary, the scenario can be changed so that both RCS and Recycler will have the second harmonic RF stations on during beam transfer resulting in a well-matched bucket-to-bucket beam transfer and smaller bunching factor. The maximum RF voltage for a cavity is chosen so that the nominal RF voltage could be achieved even if 2 of 16 (1 of 10) cavities do not operate for the first (second) harmonic RF systems, respectively.

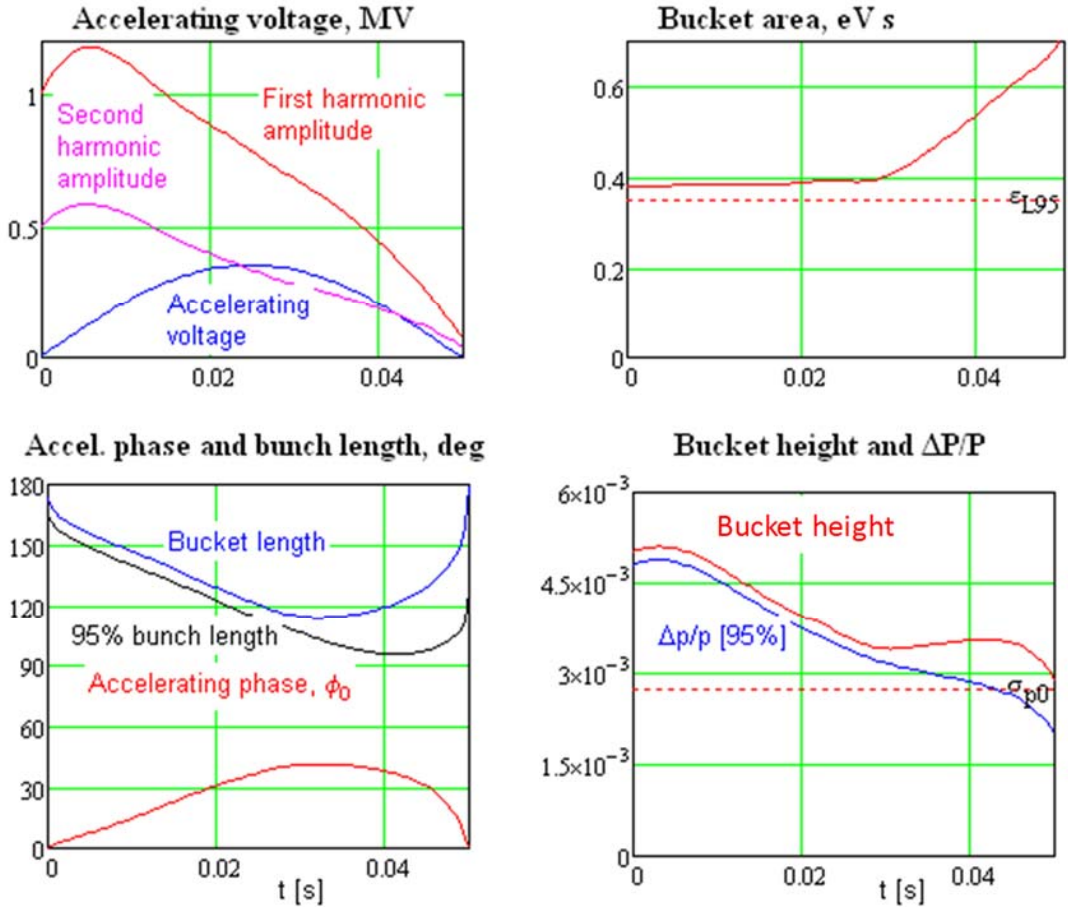

Figure 7: Beam and RF system parameters during acceleration. For this plot, the longitudinal emittance is equal to 0.35 eV s and is not changed during the acceleration.

The beam-induced voltage significantly exceeds the RF system voltage required for the beam acceleration and capture. Even at the RF voltage maximum of 1.2 MV, the beam induced voltage (on resonance) exceeds the required RF voltage by 2.6 times. This ratio achieves its maximum of about 50 at the end of the accelerating cycle where the RF voltage, depending on scenario, can achieve its minimum of about 60 kV. A relative mis-phasing of the first and second harmonic RF systems deteriorates the flat bottom potential well resulting in an increase of longitudinal density or/and shallowing the potential well and, subsequently, leading to particle losses. The required accuracy of relative phasing is about 5 deg. at the first harmonic frequency. Addressing the voltage stability and relative phasing will require a sophisticated low-level RF.

The injection to the Recycler, and, subsequently, to MI requires injection gaps of 3 buckets in the bunch structure of MI. The beam abort and the extraction from the MI at the maximum energy require an abort gap of 45 buckets in MI. Therefore, the extraction gap in RCS is chosen to be 3 buckets, same as in the present Booster.

**Table III: Main Parameters of RCS RF systems**

|  | 1-st harmonic | 2-nd harmonic |
|---|---|---|
| Harmonic number | 98 | 196 |
| Maximum voltage, MV | 1.6 | 0.7 |
| Minimum voltage, kV | 20 | 10 |
| Frequency sweep, MHz | 50.33-52.81 | 100.66 – 105.62 |
| Number of cavities | 16 | 10 |
| Shunt impedance, kΩ | 100 | 100 |

### IV.1 Longitudinal Beam Motion in a Flat-Bottom Potential Well

The equation of motion is

$$\ddot{\varphi} + \Omega_{s0}^2 F(\varphi,\varphi_0) = 0, \quad \Omega_{s0} = \omega_0 \sqrt{\frac{eV_0 q \eta}{2\pi mc^2 \gamma \beta^2}}, \tag{10}$$

where $q$ is the harmonic number, $\eta$ is the slip factor, $\omega_0$ is the revolution frequency, $e$, $m$, $\gamma$ and $\beta$ are the particle charge, mass and relativistic factors, $V_0$ is the voltage of the first harmonic RF system, and $\varphi_0$ is the accelerating phase of the first harmonic RF system, respectively. In a general case of the two harmonic RF system, the function $F(\varphi,\varphi_0)$ can be presented in the following form

$$F(\varphi,\varphi_0) = \sin\varphi - \sin\varphi_0 - v_s \sin(2(\varphi-\varphi_0)) - v_c \cos(2(\varphi-\varphi_0)) + v_c, \tag{11}$$

which was constructed so that it would satisfy $F(\varphi_0,\varphi_0) = 0$. The flat bottom potential well is determined by requirements of zeroing the first two derivatives:

$$dF(\varphi,\varphi_0)/d\varphi\big|_{\varphi=\varphi_0} = d^2 F(\varphi,\varphi_0)/d\varphi^2\big|_{\varphi=\varphi_0} = 0. \tag{12}$$

This results in

$$v_s = \frac{\cos\varphi_0}{2}, \quad v_c = \frac{\sin\varphi_0}{4}. \tag{13}$$

Consequently, the total and the accelerating voltages of the second harmonic RF system are:

$$V_1 = V_0 \sqrt{v_s^2 + v_c^2} = \frac{V_0}{4}\sqrt{1+3\cos^2\varphi_0},$$

$$V_{a1} = -V_0 v_c = -\frac{V_0}{4}\sin\varphi_0. \tag{14}$$

One can see that for the case $\varphi_0 = 0$ (no acceleration), $V_1 = V_0/2$ and both the first and the second harmonic RF systems do not transfer energy to the beam. In the case of beam acceleration, the second harmonic decelerates the beam with the decelerating voltage equal to ¼ of the voltage of the first harmonic RF system. Thus, one quarter of the power transferred to a beam from the first harmonic RF is transferred back to the second harmonic RF system.

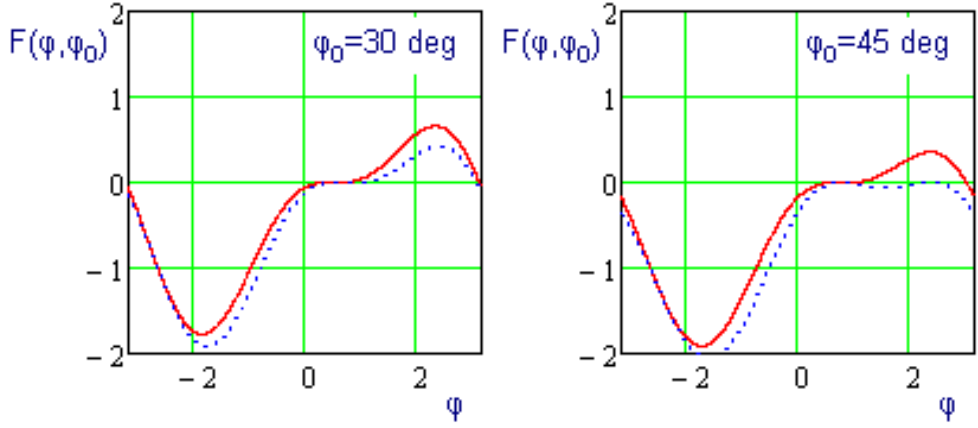

Figure 8: Dependence of $F(\varphi, \varphi_0)$ on $\varphi$ for $\varphi_0 = 30$ and 45 deg and for the cases of the flat bottom potential well (red solid line) and the case of $v_c = 0$ (blue dashed line).

To avoid the beam deceleration by the second harmonic RF system, we use the requirement of zero derivative (the first one in Eq. (12)) and $v_c = 0$. This situation is similar to the flat bottom case, which results in $v_s = \cos\varphi_0 / 2$. In this case, for small values of accelerating phase, the bottom of the well is still sufficiently flat. In contrast to the flat bottom case, such a scheme makes the bucket of finite size for $\varphi_0 < \pi/4$ only (see Figure 8)

The total area of the RF bucket is:

$$\varepsilon_L \equiv \oint p(s)ds = \frac{8mcC}{\pi q\beta}\sqrt{\frac{eV_0\gamma\beta^2}{2\pi mc^2 q\eta}}\Phi(\varphi_0). \tag{15}$$

The form factors $\Phi(\varphi_0)$ were calculated by the numerical integration for the cases of (1) the single harmonic acceleration, (2) the flat bottom well, and (3) the case of $v_s = \cos\varphi_0/2$ and $v_c = 0$. Corresponding plots are presented in Figure 9. The dependences can be also approximated by the following equations (accordingly):

$$\Phi(\varphi_0) \approx 1.145 \frac{1-\sin\varphi_0}{\left(1+\frac{1}{2}\sin\varphi_0\right)^2} \quad \text{(case 1)}, \tag{16}$$

$$\Phi(\varphi_0) \approx 1.145 \frac{\left(1-\frac{2}{\pi}\varphi_0\right)^{1.8}}{1-\frac{1}{4}\varphi_0^6} \quad \text{(case 2)}, \tag{17}$$

$$\Phi(\varphi_0) \approx 1.145\left(1 - \frac{2}{\pi}\varphi_0\right) \quad \text{(case 3)}. \tag{18}$$

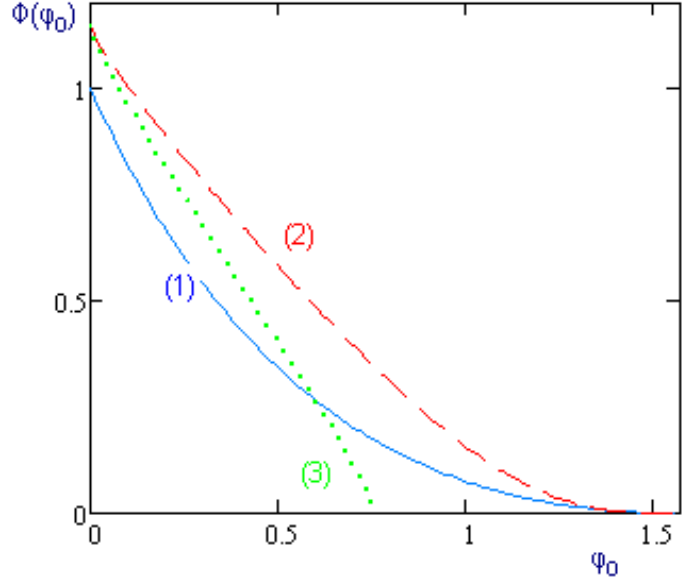

Figure 9: The form factors $\Phi(\varphi_0)$ for the cases of (1) the single harmonic acceleration, (2) the flat bottom well, and (3) the case of $v_s = \cos\varphi_0/2$ and $v_c = 0$.

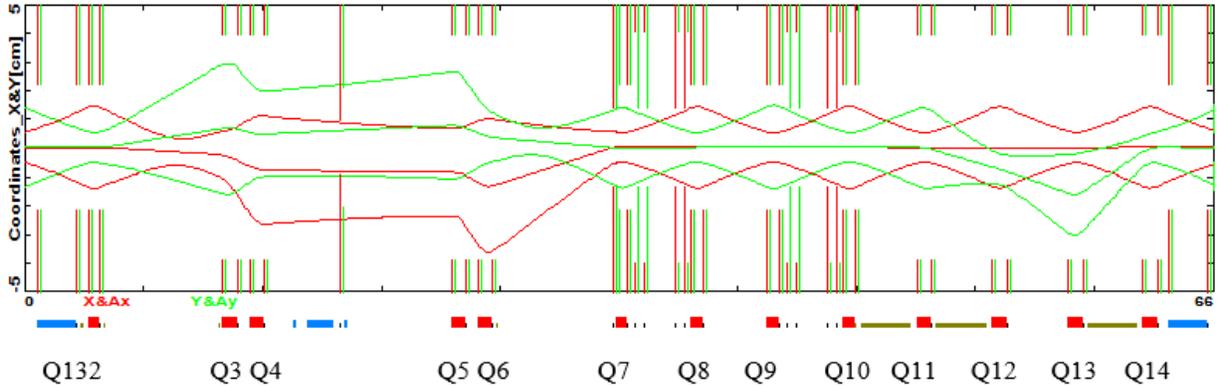

Figure 10: The beam envelopes and the aperture limitations in the injection-extraction straight line. Vertical line between Q4 and Q5 marks position of the stripping foil. Horizontal and vertical collimators are shown between quads Q7 – Q8 and Q9 – Q10. The injection bump is shown at the end of injection process after the beam is horizontally displaced from the foil.

## V  Beam injection and extraction

Figure 10 presents the beam envelopes in the injection-extraction straight line. It also shows the injection and extraction bumps and the aperture limitations. The beam envelope represents the beam boundary corresponding to the normalized emittance of 40 mm-mrad at the injection energy. The vertical lines show vacuum chamber aperture for the corresponding quadrupoles, the positions of the stripping foil, and the collimators. The injection and the extraction are on the horizontal plane from and to radially outside, respectively.

## V.1  H⁻ Multi-turn injection

The RCS will utilize a multi-turn charge-exchange injection. Due to the long injection time of 2.1 ms (~1000 turns), the transverse and the longitudinal phase space painting techniques are required to minimize the impact of the foil on the circulating beam (and vice versa) as well as to minimize the beam density for mitigation of space charge effects. The horizontal transverse injection system consists of a 3 dipole DC chicane located in the 10.43-meter straight section between the Q4 and Q5 quads. At the injection energy, the chicane displaces the reference orbit outward approximately by 17.5 mm in the middle of the merging chicane dipole. The reference orbit only moves by 6.24 mm at the foil and has a 16.4 mrad angle toward the centerline. The middle chicane dipole will be used to merge an incoming H⁻ beam with the proton orbit. Its field is limited by a loss rate due to magnetic field stripping and has been chosen to be 2.0 kG resulting in particle losses of $1.3 \cdot 10^{-6}$ ($8.7 \cdot 10^{-7}$ m⁻¹).  The field in the third chicane was chosen to be 8.333 kG such that it will strip any H⁻ emerging from the foil or missing it. It will also strip higher excited states of H⁰ coming out from the foil. Most of them will be accepted to the beam but some, which penetrated deep into B3 field before stripping, will be lost in the collimation system located immediately downstream of the injection region.  To reduce the angular spread due to stripping in the magnetic field, the field edge of the dipole has to be short. For symmetry, the first chicane dipole has the same field as the third. Figure 11 presents layout of the injection region. Magnetic fields of the chicane dipoles are not changed during acceleration. They excite ±5% betatron wave on the vertical plane that, if required, can be easily corrected by trim quadrupoles.

The four fast correctors (on each plane) perform the transverse beam painting. Such a scheme allows one to have the independent control of the closed orbit position and the angle for both planes at the foil location. The correctors are located near the quads 132, 3, 6 & 7. The orbit bump consists of two constituents: the painting bump (changing during injection) and the injection bump designed to minimize the required aperture in the injection region quadrupoles. After painting the beam is moved out of the foil horizontally to inward and is parked at -7.61 mm relative to the reference orbit. With further acceleration the value of this displacement is decreasing together with the beam size resulting in that the beam does not touch the foil even if values of the injection bump correctors stay unchanged during acceleration. The injected H⁻ beam does not move during injection. Relative to the reference orbit (which takes into account the beam displacements in the injection dogleg) it has the following coordinates on the foil: $x = 11.80$ mm, $y = 5.85$ mm, $\theta_x = -0.527$ mrad and $\theta_y = 0.284$ mrad. The coordinates of the foil corner are $x_f = 9.25$ mm, $y_f = 3.38$ mm. Figures 12 and 13 show the beam and vacuum chamber cross-sections on the foil and the beam envelopes in the injection region. To have enough room for the injection bumps, the aperture of four quads (Q3-Q6) is increased from 25 to 45 mm.

The foil thickness has been chosen to be 600 µg/cm². This is thick enough to strip the major fraction of H⁻ beam to protons leaving about 0.5% particles as H⁰ and negligible fraction as H⁻. Further increase of the thickness would result in larger particle losses due to scattering in the foil and larger foil heating. However, both these problems stay on the manageable level (see below). To minimize the number of secondary passages through the foil, the beam is parked quite close to the foil edges (2.32σ in both planes) resulting in that about 2% particles will be missing the foil. They will be converted to H⁰ in the field of B3 and go to the beam dump. The finite length of B3 edge field increases the angular spread of field stripped H⁻. The RMS width of this angular distribution is about 1 mrad which is about 5 times larger than the RMS horizontal angular spread of the H⁰ beam. The center of the distribution is shifted by ~3 mrad. Figure 14 shows the distribution of H⁰ beam after stripping of H⁻ in the field of B3 dipole.

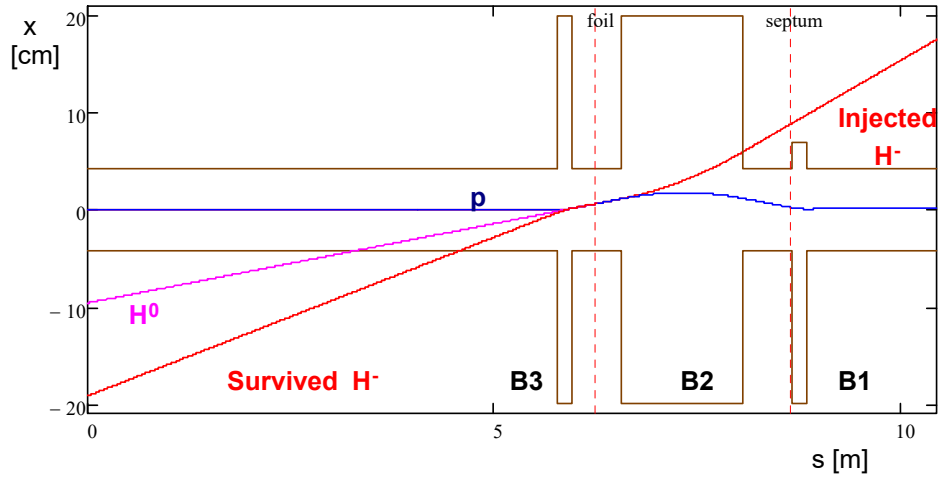

Figure 11: The trajectories of the circulating proton and incoming H⁻ beams. The brown lines show aperture of vacuum chamber and positions of chicane dipoles.

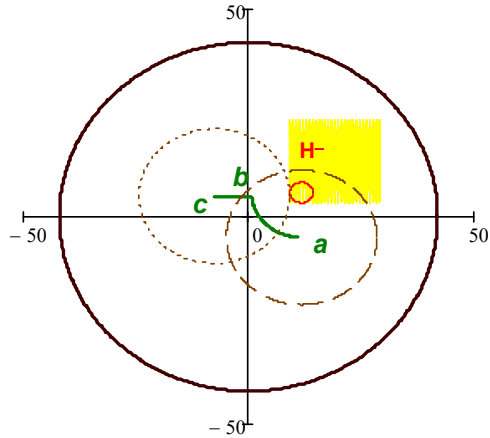

Figure 12: The beam and vacuum chamber cross-sections at the foil. The green line shows the displacement of the closed orbit during painting. It starts at point *b*, moves to point *a*, then goes back to *b*, and finally it is moved to point *c* to prevent further beam interaction with the foil. The yellow square shows position of the stripping foil. The red ellipse shows the boundary of an injected H⁻ beam. The brown dashed and dotted lines present the boundary of a stored proton beam when the closed orbit is located in points *a* and *c* for the machine acceptance of 40 mm mrad (normalized). The internal radius of the vacuum chamber is 42 mm.

The total power of the injected beam is about 85 kW. About 4% of these particles are lost during injection: ~2% miss the foil, 0.5% are not completely stripped in the foil, 0.15% are single scattered in the foil, and ~1% are outside of 40 mm mrad RCS acceptance and are lost in the scraping system. In the normal operating conditions, it results in the heat load on the injection beam dump being about 3 kW and 1.5 kW intercepted by the collimation system. A prudent design (as confirmed by the SNS experience) would have both the injection waste beam absorber and the collimation system designed to handle about 10% or 8.5 kW.

Stripping of H⁻ also yields two 1.1 MeV electrons per each stripped H⁻. These electrons carry sufficiently large power of about 90 W that needs to be intercepted by an electron beam dump. After leaving the foil, the electrons are reflected from the B3 dipole where they are bent by its magnetic field. It results in the bending angle of 180 deg (reflection) and the beam shift of about -9 cm on the horizontal plane. That minimizes interference between the stripping system and the electron dump. The B3 field also results in a vertical defocusing of the electron beam and, consequently, a reduction of peak heat load. A design of the electron dump has to prevent the interaction of secondary electrons with the circulating beam.

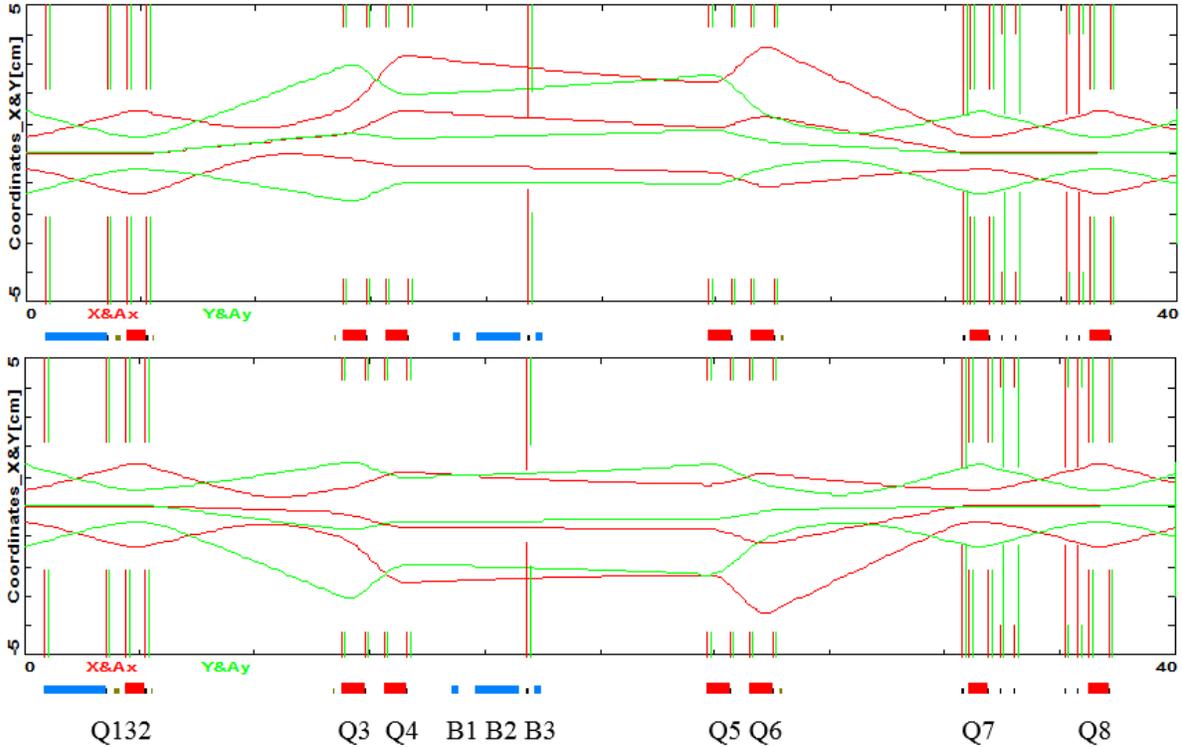

Figure 13: Beam envelopes through the injection bump region for parameters of Figure 2; top – injection bump only, bottom – injection bump + maximum amplitude for X&Y painting bumps. Vertical green line near Q7 shows primary horizontal scraper.

## V.2 Transverse and longitudinal injection painting

We considered a few possible transverse painting schemes. In this document we present a scheme that minimizes the foil heating, is sufficiently simple, and has the beam distribution close to the desired KV-distribution. The above infrastructure of the injection region is sufficiently flexible and does not prevent us from using other painting methods in the future.

The following criteria were used to create a particle distribution: (1) create a KV-like distribution which would have approximately a constant particle density across the beam, (2) make the beam cross section on the *x-y* plane to be elliptical with semi-axes corresponding to horizontal and vertical emittances of 22 mm-mrad, and (3) minimize the number of secondary passages through the foil. The time dependence of closed orbit position on the foil for the first half of the painting cycle can be described by the following equations,

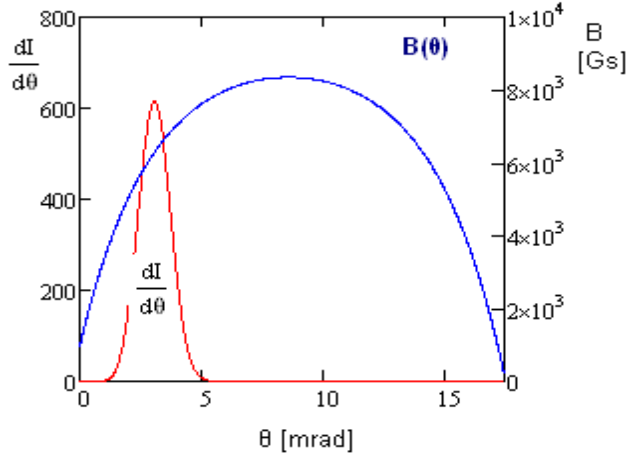

Figure 14: Angular distribution of $H^0$ after $H^-$ stripping in field of B3 dipole (red line). Blue line presents the magnetic field in the corresponding stripping point.

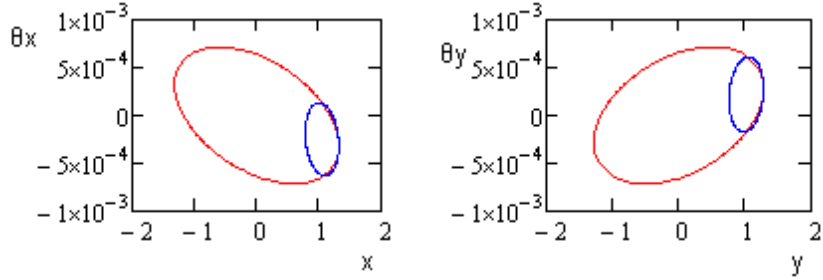

Figure 14: Phase space boundaries of the linac (blue lines) and RCS (red lines) beams. The linac beam boundaries correspond to the normalized boundary emittance of 3 mm mrad (2.45σ of 0.5 mm mrad RMS emittance) the RCS beam boundaries correspond to the normalized emittances of 25 mm-mrad.

$$A_x(t) = -A_{x0} \cos(\phi(t)), \quad \phi(t) = \begin{cases} \phi_0 + 2(\pi/2 - 2\phi_0) t/T_{inj}, & t \leq T_{inj}/2, \\ \phi_0 + 2(\pi/2 - 2\phi_0)(T_{inj} - t)/T_{inj}, & t \leq T_{inj}/2, \end{cases} \quad (19)$$
$$A_y(t) = -A_{y0} \sin(\phi(t)),$$

where $A_{x0}$ = 11.1 mm, $A_{y0}$ = 10.8 mm, $\phi_0$ = 79 mrad, $T_{inj}$ = 4.3 ms. In the second half of the painting cycle, the closed orbit motion is inverted so that it comes to the initial point at the cycle end. To minimize the number of passages through the foil, all beta- and alpha-functions of the linac beam are scaled from the corresponding values of RCS by factor of 0.345 so that the linac phase space would be inscribed into the $x$ and $y$ machine acceptances as shown in Figure 16. To paint on the KV-distribution (in the 4D phase space), the angles and positions of the injected linac beam are scaled together so that:

$$\theta_x = x/L_x,$$
$$\theta_y = y/L_y, \quad (20)$$

with $L_x$ = -42 m and $L_y$ = 46 m. Such correlation also reduces the number of passages through the foil. All these measures resulted in the distribution quite close the KV-one. Figure 16 presents particle positions at the end of painting. One can see that the almost all particles are inside the ellipse corresponding to the horizontal and vertical normalized emittances of 35 mm mrad. Figure 17 presents the single dimensional particle densities for horizontal and vertical planes at the foil location and their comparison to the KV distribution with for $\varepsilon_{xn}=\varepsilon_{yn}$=22 mm mrad. One can see a good coincidence for the main part of the distributions. The distribution tails are related to the finite value of linac emittances. Figure 18 presents the integral particle distributions over 2D and 4D Courant-Snyder invariants:

$$I_x = \theta_x^2 \beta_x + 2\alpha_x x \theta_x + (1+\alpha_x^2) x^2 / \beta_x,$$
$$I_y = \theta_y^2 \beta_y + 2\alpha_y y \theta_y + (1+\alpha_y^2) y^2 / \beta_y, \qquad (21)$$
$$I_{4D} = I_x + I_y \ .$$

The total number of particles outside of machine acceptance of 40 mm mrad (normalized) is about 1%. The initial optimization and tracking were done neglecting the beam space charge effects. Tracking studies with the beam space charge taken into account were done using the ORBIT code. They exhibited only a minor effect of beam space charge on the final distribution and particle losses.

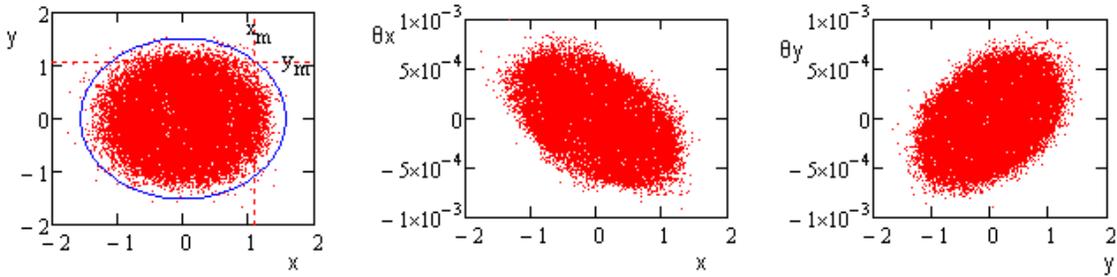

Figure 16: Particle coordinates at the end of painting. Blue line is shown for the horizontal and vertical normalized emittances of 35 mm mrad.

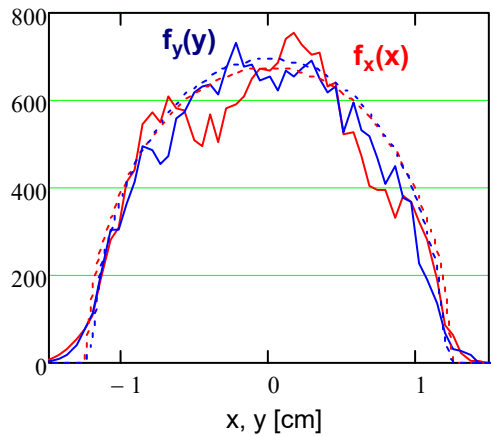

Figure 17: Dependence of single dimensional particle densities on transverse coordinates at the foil location. The dotted lines show the corresponding densities for the KV-distribution with $\varepsilon_{xn}=\varepsilon_{yn}$=22 mm mrad.

The simulations exhibited that the number of secondary passages through the foil is about 50 per incoming particle. Figure 19 presents the distribution of secondary passages through the foil surface. The use of double pass painting, described above, resulted in a more uniform power distribution on the foil and, corresponding, reduction of peak power density by ~20% relative to a single pass painting. The maximum of the distribution is achieved at the foil corner and is equal to ~2.2 passages per mm2 per incoming particle. The density of the incoming H⁻ beam is 0.14 mm-2 per particle. It is shifted from the foil edges and does not change the maximum of power deposition located at the foil corner. To improve the foil cooling, the foil of 420 µg/cm2 thickness is rolled by 45 deg. so that it would have the stripping power of 600 µg/cm2 foil. Taking into account that about 25% of heating power is removed by δ-electrons and that the foil is cooled by the black body radiation on its both sides, one obtains that the foil peak temperature is ~1200 Co.

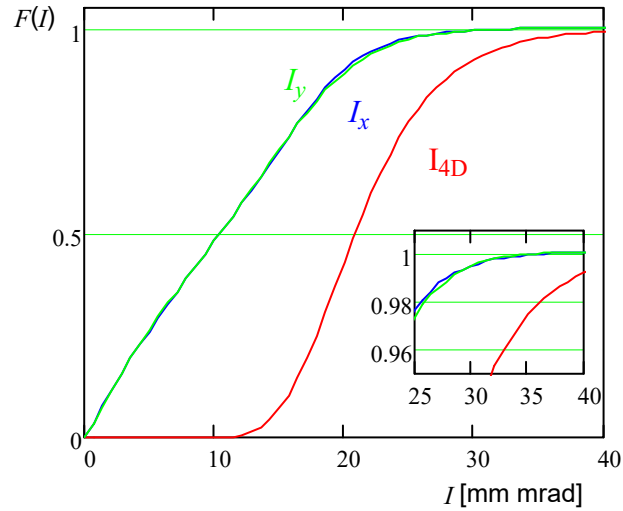

Figure 18: The integral particle distributions over Courant-Snyder invariants. Inset shows the upper end of the same curves.

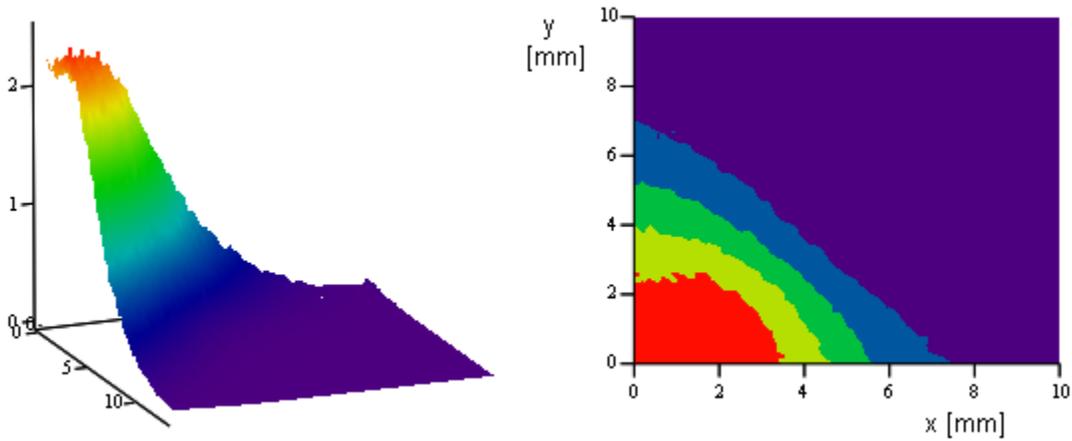

Figure 19: Distribution of secondary passages on the foil surface; the peak of the distribution corresponds to 22 passages per mm² per injected particle.

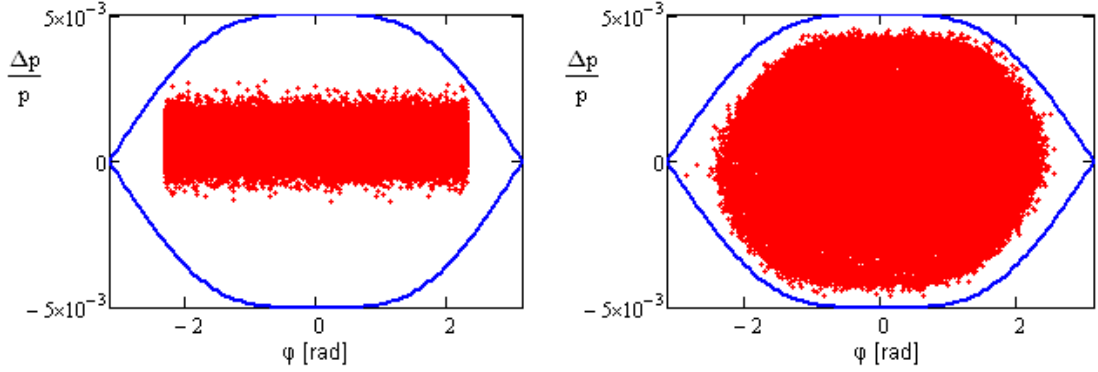

Figure 20: The phase space positions of the injected particles (left) and the particle positions at the end of injection ($\sigma_p=5\cdot 10^{-4}$, $\Delta p/p=7\cdot 10^{-4}$, $T_w=14.6$ ns (73%))

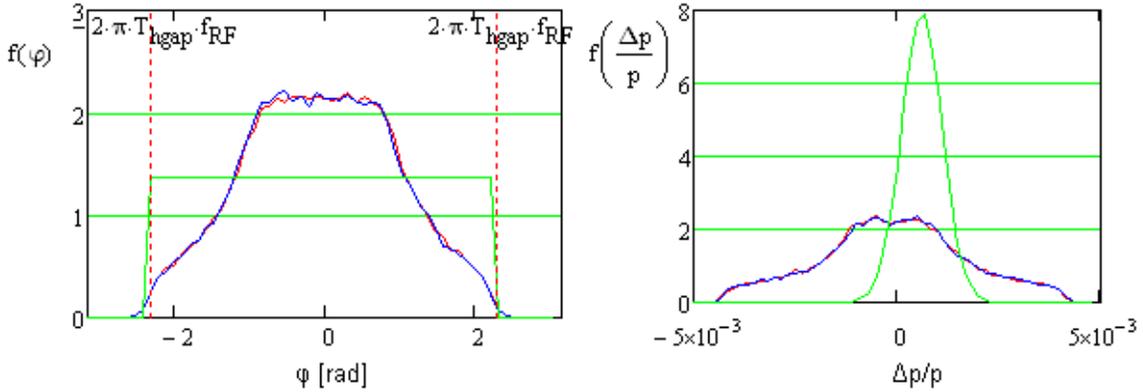

Figure 21: Particle distributions over RF bucket length (left) and over momentum (right); green lines – for injected particles, red lines – after injection, blue lines - 500 turns after injection.

The painting of the longitudinal distribution is performed by the momentum offset of linac beam relative to the RCS RF bucket center. The dual harmonic RF system is used to flatten the bucket. The nominal voltage of 1 MV (the first harmonic) is on at the injection beginning. The value of the offset is $\Delta p/p = 7\cdot 10^{-4}$ while the bucket height is $5\cdot 10^{-3}$. To prevent the particle injection outside of the RF buckets, the linac bunches which can inject such outside particles are chopped off. The chopper is synchronized with the RCS RF and chops off particles located outside ±7.3 ns window (relative to the bucket center) corresponding to 73% of the bucket length. Simulation results are shown in Figures 20 and 21. The bunch is sufficiently flat at the top of the distribution. Its bunching factor is equal to 2.2. The contribution of the beam space charge and the vacuum chamber impedance are small and were neglected in these simulations.

### V.3 Beam extraction

The beam extraction from the RCS is located in the downstream end of the injection/extraction straight section as shown in Figure 22. The extraction plane is vertical, where the beam is kicked vertically into a horizontally bending Lambertson magnet, similar to the configuration utilized in other Fermilab accelerators. There are two kickers located between the quads 10 and 12. The Lambertson magnet is located between the quads 13 and 14. This configuration will be responsible for all extractions from the RCS.

In normal operating conditions, the RCS extraction energy is matched to the Recycler energy that cannot be changed because the Recycler is permanent-magnet fixed-energy ring. The initial configuration does not include an RCS beam abort system` to dump the beam during acceleration. In the case of an incident, the beam energy will be absorbed by the beam collimation system, which is not designed to withstand permanent loss of full intensity beam but can withstand infrequent loss of a single full intensity pulse with stored energy of 34 kJ.

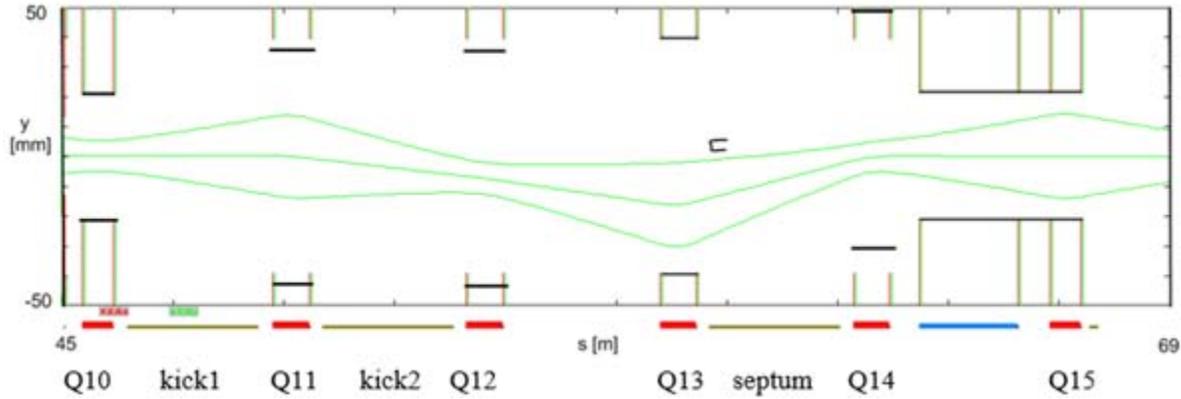

Figure 22: Vertical beam envelope (at $40\pi$) of injected beam in the extraction region. Beam displacement at septum is 14.3 mm. Horizontal black lines show aperture in displaced quads: Q11 = -4.8 mm, Q12 = -6.39 mm, Q14 = 9.84 mm.

The vertical orbit displacement below the Lambertson septum is achieved by displacement of 3 quadrupoles (Q11, Q12 and Q14) making a permanent extraction bump. Figure 22 shows the vertical beam envelope of injected beam with its displacement taken into account. It also shows the position of the septum. The Lambertson septum is assumed to be 5 mm thick and a minimum of 5 mm is left between the septum and circulating beam at all energies. To have enough room, 4 quads (Q11 – Q14) have increased aperture of 45 mm (radius). They have the same design as the injection quads.

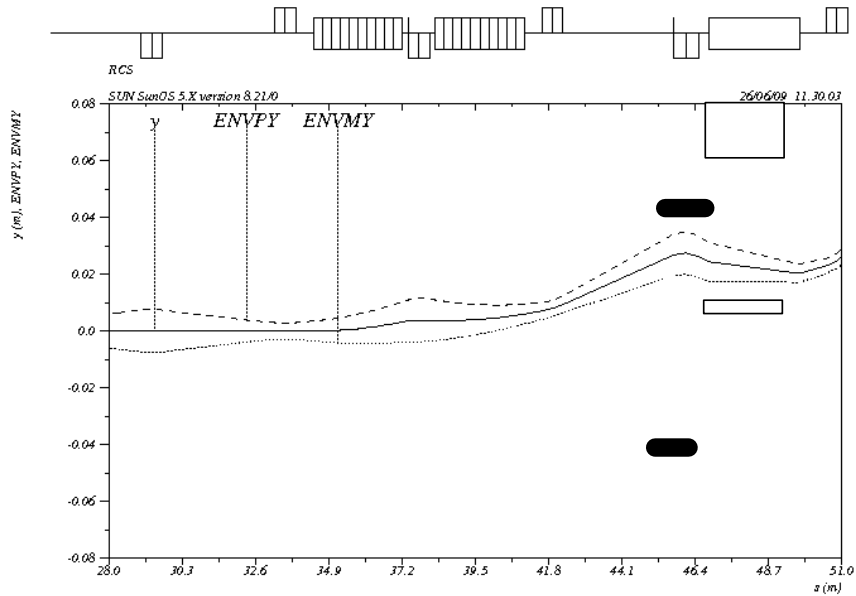

Figure 23: An 8-GeV extraction beam envelope through the Lambertson field region.

Figure 23 shows the extracted beam orbit and 8 GeV beam envelope due to the ~4.5 mrad vertical kick of the extraction kicker pair. This produces a 24 mm displacement at the entrance of the Lambertson with a 6 mrad angle toward the centerline. The horizontal Lambertson has been rolled by ~7 degrees to compensate the downward kick from the quadrupole just upstream of the Lambertson. For the initial layout, the Lambertson horizontal bend angle has been chosen to be 96 mrad to be able to install the first quadrupole magnet of the extraction line in the space between the RCS quad and dipole. This angle requires a dipole field of approximately 1 Tesla for magnet length of 2.85 meters. The vertical offset of quad upstream of the Lambertson or its aperture needs further evaluation. The aperture and magnet choices of this configuration will be re-evaluated if this system will need to be utilized as a RCS abort system.

### V.4 Injection Beam Dump and Beam Collimators in RCS

The injection dump and the collimation system localize particle losses. They are well protected from the radiation excited by the beam. It results in a significant reduction of the residual radiation in the RCS tunnel. The injection dump is installed immediately downstream of the injection chicane (before the quad Q5). It intercepts $H^0$ which are not stripped to protons. The dump design allows routine operation with 4 kW beam loss and can be operated up to 10 kW during limited time (hours).

The closest existing collimation systems are those of SNS and JPARC. The JPARC accelerator complex has a 3-GeV proton ring to operate at ~1 MW. It has two primary (1-mm tungsten foils) and seven massive secondary collimators (iron). The initial design of the RCS collimation system suggests that it will consist of a primary momentum collimator, primary betatron collimators and 4 major collimators. The primary momentum collimator is located downstream of the quad Q130, it has 2 jaws allowing to scrape the beam horizontally on both sides. The horizontal and vertical primary betatron collimators are located downstream of the quad Q7. Each of them scrapes the beam on the two sides. The four main collimators (2 horizontal and 2 vertical) are located between the quads Q8 and Q10, so that each half cell would have one horizontal and one vertical collimator. Each collimator has two jaws and scrapes both sides of the beam. The jaws of each main collimator have an independent control of their angle and position so that they could be aligned along the beam envelope.

The RCS injection beam dump design is based on the existing devices, already in operation in the Fermilab accelerator complex. The design is based on iron and/or tungsten jaws which encounter the beam first. The radiation shielding is accomplished with thick radial and axial outer iron plates, surrounded by marble and/or concrete. Cooling is performed with borated water (2000 parts of $^{10}B$ per $10^6$) that removes several tens of kW. Energy deposition calculations for the injection absorber were performed with the MARS simulation code. The absorber is designed to localize the beam loss at injection and to meet the requirements of the Fermilab Radiological Control Manual (FRCM) [6]. The following issues were addressed: (i) surface water activation; (ii) residual activation; (iii) survival of the magnets around the absorber; (iv) cooling. Current design of the absorber with an inner radiation trap is shown in Figure 24. The beam intensity is $2.67 \cdot 10^{13}$ 2-GeV proton/pulse at the repetition rate of 10 Hz, which corresponds to the beam power of 85 kW. It is assumed that 4% of the beam is deposited in the absorber. The beam pipe radius is equal to 4 cm.

According to the Fermilab Concentration Model, the activation of the surface water for a given beam intensity reaches its permitted maximum approximately every 6 months, so that the collected water must be replaced twice a year. Calculated residual doses around the absorber are shown in Figure 25. There are a couple of hot spots on the tunnel wall and side surface of the quadrupole with residual dose of about 200 and 700 mRem/hr, respectively, while typical requirement for hands-on maintenance is 100 mRem/hr. The issue can be resolved by means of extra marble shielding layers. The power density was calculated for the coils of the 1st magnet both upstream and downstream of the absorber. The power density in the 1st magnet upstream is pretty low, so that it can survive for about 100 years. The 1st quadrupole downstream of the absorber can survive for about 8 years. In the estimate it is assumed that, according to empirical observations, the lifetime of a magnet depends on the lifetime of epoxy resin used in its coils and the latter corresponds to absorbed dose of 400 Mrad.

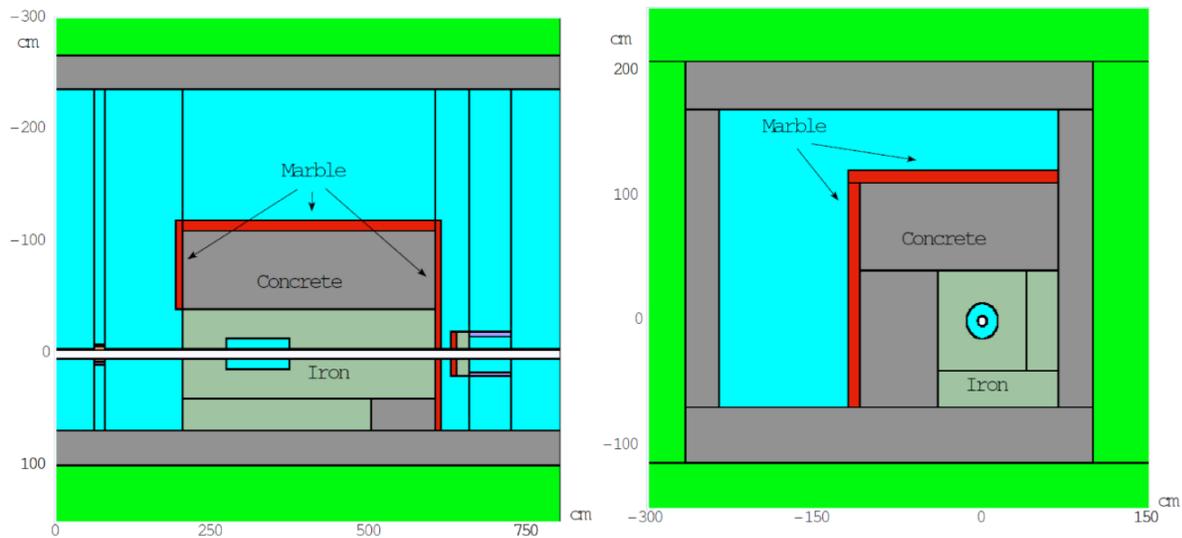

Figure 24: Plan view (left) and cross section (right) of the injection absorber in the tunnel. The following color coding is employed: white, blue, green, light gray, dark gray, red and violet correspond to vacuum, air, soil, iron, concrete, marble and yoke, respectively.

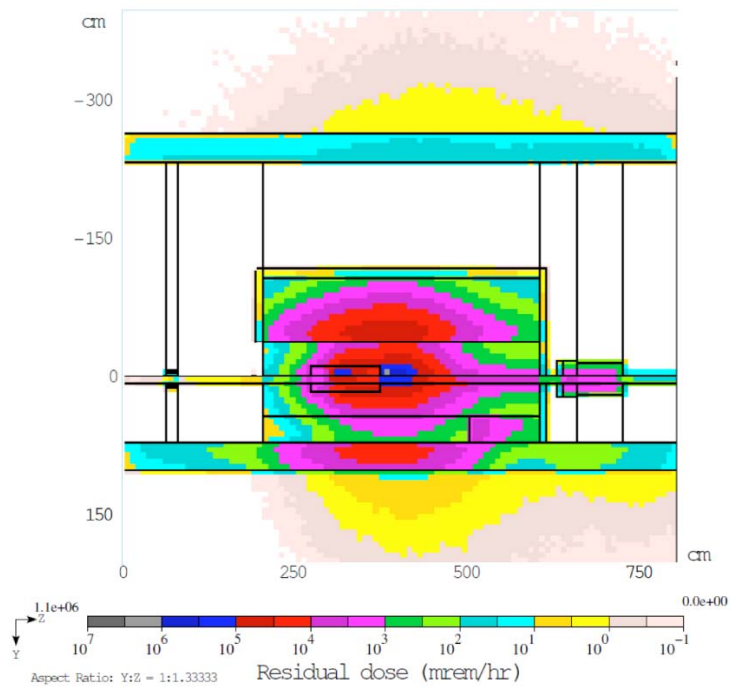

Figure 25: The calculated residual dose distribution (plan view) around the injection absorber for standard irradiation conditions: 30-day irradiation followed by a 1-day cooling.

Finally, passive cooling of the absorber with airflow appears to be adequate. Using the passive cooling rate of about $10^{-3}$ W/(cm$^2$ K) and the outer surface of the absorber of $2.2 \cdot 10^5$ cm$^2$, one can see that the absorber's temperature will exceed the temperature of the surroundings by approximately 10 K.

The length of the absorber is determined mostly by the requirement to keep the absorbed dose in the coils of the 1$^{st}$ quadrupole downstream of the absorber at an acceptable level. The transverse dimensions depend mostly on residual activation and surface water activation. The total weight of the absorber is about 58 tons including 32 tons of iron, 20 tons of concrete and 5.6 tons of marble. Further optimization of the design can be performed.

## V.5 RCS Magnets

In the previous Fermilab studies of 16 GeV and 8 GeV Booster replacements, the dipole magnets had large good-field areas: 127 mm x 228.6 mm in Study I [7], and 101.6 mm x 152.5 mm in Study II [8]. In the present RCS concept, the beam aperture is substantially reduced to ~40 mm diameter which leads to smaller magnets, lower stored energy and operational expenses.

The RCS optics is based on a separated function lattice and magnets: 100 ring dipoles and 132 quadrupoles are connected in series and powered by a resonant 10-Hz power supply. Magnet yokes are laminated and made from M17 coated low carbon steel.

The RCS dipole is an iron dominated magnet and has a conventional H-magnet design. A proper pole tip shimming with minimization of pole tip width and peak field optimizes field homogeneity. The magnet cross-section is plotted in Figure 26 and the parameters are shown in Table IV.

The RCS quadrupole magnet parameters are shown in Table V and its cross-section is shown in Figure 27. The quadrupole integrated strength will be corrected by ± 2.6 % (±0.3 T) by a shunting power supply or an additional trim winding wound over main coils or trim quads.

Multipole correctors will have vertical or horizontal dipole, and sextupole windings. The corrector parameters are shown in Table VI. The magnet design and the fabrication technology are similar to that of the existing Fermilab Booster correctors [9], [10] which are in operation now. The number of poles is reduced from 12 to 6, which is enough for the combined dipole and sextupole correctors. In this case the horizontal corrector (vertical field) and vertical corrector (horizontal field) have different winding configurations to explore the six-pole yoke geometry. The quadrupole correction combined with the main quadrupole by using trim coils or shunting power supplies. The corrector magnet has two modifications: horizontal corrector with sextupole and vertical corrector with sextupole. The quantity and corrector positions in the lattice are described in Section II.

The correctors have laminated yokes that are split in the vertical direction and have indirect water cooling. The laminations are coated to reduce the eddy current effects and provide proper half yokes curing. The coils could be randomly wound into slots between the poles around the yoke. Finally, the magnet is assembled from two sets into a single solid package impregnated with epoxy.

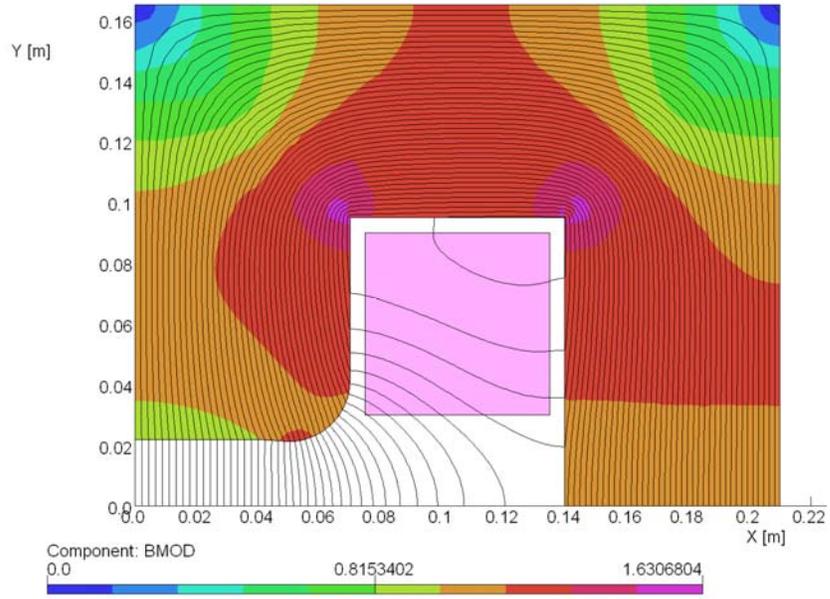

Figure 26: Dipole magnet geometry with flux lines and flux density at 667 A current.

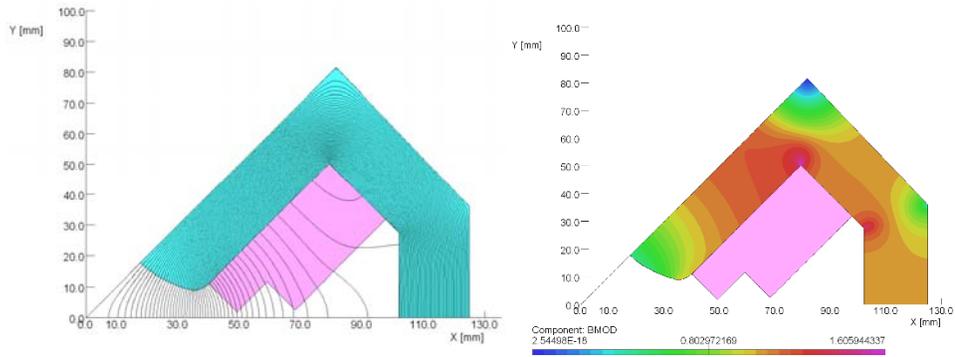

Figure 27: The RCS quadrupole geometry, flux lines (left), and the yoke flux density(right) at 672 A current.

Table IV:  Parameters of RCS dipole magnets

| Parameter | Unit | Value |
|---|---|---|
| Number of magnets |  | 100 |
| Peak field | T | 0.87375 |
| Field at injection | T | 0.2184 |
| Magnet gap | mm | 44 |
| Good field area diameter | mm | 40 |
| Field homogeneity |  | 0.02 % |
| Effective length | m | 2.13216 |
| Peak current | A | 667 A |
| Current form |  | Offset sine wave |
| Current frequency | Hz | 10 |
| Duty factor | % | 100 |
| Number of turns/pole |  | 24 |
| Copper conductor | mm x mm | 12.5 x 12.5 |
| Conductor cooling hole | mm | 7 |
| Number of pancake coils/pole |  | 2 |
| Lamination material |  | M17 |
| Lamination thickness | mm | 0.35 |
| Inductance | H | 0.025 |
| DC resistance | Ohm | 0.021 |
| Stored energy | kJ | 5.47 |
| Power losses RMS (without | kW | 4.3 |
| Peak inductive voltage | V | 390 |
| Number of cooling |  | 1 |
| Water pressure drop | MPa | 0.5 |
| Water flow | l/min | 2.8 |
| Water temperature rise | Cº | 22 |

Table V: Parameters of RCS quadrupoles

| Parameter | Unit | Value |
|---|---|---|
| Peak field gradient | T/m | 17.43 |
| Field gradient at injection | T/m | 5.47 |
| Pole tip radius | mm | 25 |
| Good field area diameter | mm | 40 |
| Field nonlinearity (2D) | | 0.03 % |
| Effective length | m | 0.612 |
| Peak current | A | 672 A |
| Current form | | Offset sine wave |
| Current frequency | Hz | 10 |
| Duty factor | % | 100 |
| Number of turns/pole | | 7 |
| Copper conductor | mm x mm | 10 x 10 |
| Conductor cooling hole | mm | 5 |
| Number of coils/pole | | 1 |
| Lamination material | | M17 |
| Lamination thickness | mm | 0.35 |
| Inductance | mH | 1.15 |
| DC resistance | Ohm | 0.012 |
| Stored energy | J | 260 |
| Power losses RMS (without | kW | 2.0 |
| Peak voltage | V | 40 |
| Number of cooling | | 1 |
| Water pressure drop | Mpa | 0.5 |
| Water flow | l/min | 1.9 |
| Water temperature rise | Cº | 16 |

**Table VI: Parameters of RCS corrector magnets**

| Parameter | Unit | Vertical | Horizontal | Sextupole |
|---|---|---|---|---|
| Number of magnets |  | 66 | 66 | 132 |
| Peak dipole field | T | 0.047 | 0.054 |  |
| Peak sextupole gradient | T/m$^2$ |  |  | 200 |
| Magnet aperture | mm | 44 | 44 | 44 |
| Effective length | m | 0.2 | 0.2 | 0.2 |
| Peak current | A | 10 | 10 | 10 |
| Current form |  | Offset sine | Offset sine | Offset |
| Current frequency | Hz | 10 | 10 | 10 |
| Duty factor | % | 100 | 100 | 100 |
| Number of turns/pole |  | 90 | 90 | 50 |
| Copper conductor area | mm$^2$ | 3 | 3 | 3 |
| Lamination material |  | M17 | M17 | M17 |
| Lamination thickness | mm | 0.35 | 0.35 | 0.35 |
| Inductance | H | 0.03 | 0.03 | 0.018 |
| DC resistance | Ohm | 1.2 | 1.2 | 1.0 |
| Stored energy | J | 1.5 | 1.5 | 0.9 |
| Power losses RMS | W | 60 | 60 | 50 |
| Peak inductive voltage | V | 20 | 20 | 12 |
| Water cooling |  | indirect | indirect | indirect |

# VI Summary

We have described the cost-effective 2-8 GeV Rapid Cycling Synchrotron, capable of delivering 340-kW beam power at 10 Hz and to support the Fermilab Main Injector 2-MW operations. The cost-effectiveness is achieved by reducing the size of the vacuum chamber, which in turn, allows for more compact magnets and less complex rf cavities. While such a concept supports the 2-MW Fermilab MI beam power, it is limited at < 500 kW of the RCS beam power because of potential space-charge and injection limitations.

# VII Acknowledgements

We would like to thank many contributors to this article, including Steve Holmes, Paul Derwent, Alexey Burov, Nikolay Mokhov, Igor Rakhno, Vladimir Kashikhin, Brian Chase, and John Reid. Fermilab is Operated by Fermi Research Alliance, LLC under Contract No. DE-AC02-07CH11359 with the U.S. Department of Energy.